%% file: Main.tex
\documentclass[%
 reprint,floatfix,
nofootinbib,
 amsmath,amssymb,
 aps,
 prl,
]{revtex4-2}

\usepackage{graphicx}
\usepackage{dcolumn}
\usepackage{bm}
\usepackage{standalone}
\usepackage{import}

\usepackage[english,main=english]{babel}

\usepackage{siunitx}
\let\origsi\si
\renewcommand{\si}{\,\origsi}

\newcommand{\BB}[0]{K$_{0.3}$MoO$_3\,$}
\begin{document}
\preprint{APS/123-QED}

\title{Ultrafast Decoherence of Charge Density Waves in K$_{0.3}$MoO$_3$}

\author{Rafael T. Winkler$^{1, 4}$, Larissa Boie$^1$, Yunpei Deng$^2$, Matteo Savoini$^1$, Serhane Zerdane$^2$, Abhishek Nag$^2$, Sabina Gurung$^1$, Davide Soranzio$^1$, Tim Suter$^1$, Vladimir Ovuka$^1$, Janine Zemp$^1$, Elsa Abreu$^1$, Simone Biasco$^1$, Roman Mankowsky$^2$, Edwin J. Divall$^2$, Alexander R. Oggenfuss$^2$, Mathias Sander$^2$, Christopher Arrell$^2$, Danylo Babich$^2$, Henrik T. Lemke$^2$, Urs Staub$^2$, Jure Demsar$^3$, Steven L. Johnson$^{1,2}$ }
\affiliation{$^1$Institute for Quantum Electronics, Physics Department, ETH Zurich, Zurich, Switzerland. \\
$^2$Center for Photon Science, Paul Scherrer Institute, Villigen, Switzerland. \\
$^3$Institute of Physics, Johannes Gutenberg-University Mainz\\
$^4$Institute of Science and Technology Austria (ISTA), Am Campus 1, 3400 Klosterneuburg, Austria
}
\date{\today}

\begin{abstract}
Recent works have suggested that transient suppression of a charge density wave (CDW) by an ultra-short excitation can lead to an inversion of the CDW phase. We experimentally investigate the dynamics of the CDW in \BB by time resolved x-ray diffraction after excitation with optical pulses. Our results indicate a transient inversion of the CDW phase close to the surface that evolves into a highly disordered state in less than one picosecond. Numerical simulations solving the Ginzburg-Landau equation including disorder from strong pinning defects reproduce our main observations. Our findings highlight the critical role of disorder in schemes for coherent control in condensed matter systems.
\end{abstract}

\maketitle

Attempts to manipulate order parameters in solids using intense femtosecond light sources have been a strong focus of research in condensed matter physics over the last couple of decades~\cite{Torre_2021, Kirilyuk_2010, Disa_2021}. Despite many fascinating observations of control over different kinds of order, a general understanding remains incomplete. The effects of spatial disorder in particular remain poorly understood, although some recent experiments indicate its importance in constraining efforts towards coherent control of the order parameter~\cite{Wall_2018}. 

Charge density waves (CDW) are observed in many solid state systems at low temperatures, arising from strong coupling of electronic states near the Fermi level to a set of phonon modes. Several recent studies have provided evidence for coherent control over topological defects in CDWs~\cite{Yusupov_2010, Trigo_2021}. The basic concept is that an ultrashort laser pulse deforms the free energy landscape from its broken symmetry ground state to a high symmetry potential which then relaxes back on time scales comparable to the oscillation periods of the phonon modes involved. For certain values of the excitation density, coherent oscillations of the density wave in the high-symmetry phase can cause a reversal of the phase of the restored CDW relative to the original order. This behavior is in principle not limited to CDW materials and could be extended to other systems undergoing second-order phase transitions. 

Here we report time-resolved x-ray diffraction of these dynamics in \BB, a quasi 1-D material 
 with a prototypical incommensurate CDW with wave vector $2k_F =$ (1, 0.748, 0.5) \cite{Schutte_1993}. 
 To numerically reproduce the experimental observations, we model the dynamics with a modified Ginzburg-Landau equation. We find that reproducing the experimentally observed dynamics requires a two-dimensional (i.e. complex) order parameter and consideration of three spatial dimensions and, crucially, the inclusion of strong pinning defects.

The time-resolved x-ray diffraction experiments were performed at the SwissFEL free electron laser at the Paul Scherrer Institute \cite{Prat_2020, Ingold_2019, Mankowsky_2021}. The sample was excited by a $30\si{\femto\second}$ laser pulse at a wavelength $\lambda = 510 \si{\nano\meter}$, generated by optical parametric amplification pumped by an amplified Ti:Sapphire laser system. The excitation wavelength is chosen to maximize the optical intensity attenuation length $\delta=190\si{\nano\meter}$ \cite{Sagar_2007}  to ensure matching with the $210\si{\nano \meter}$ intensity attenuation length of the x-rays at a grazing incidence angle $\alpha = 0.73^\circ$ \cite{Henke_1993}. The dynamics of the periodic lattice distortion (PLD) induced by the CDW was probed by time resolved x-ray diffraction from the (3, 7.252, -2.5) super-lattice peak with x-ray pulses of $50\si{\femto\second}$ duration and a photon energy of $7.5\si{\kilo\electronvolt}$. The diffracted x-ray intensity, a direct measure of the amplitude of the CDW lattice distortion, was measured by a 2D pixel detector (Jungfrau \cite{Redford_2016}). By rotating and scanning the sample by an angle \(\phi\) about its surface normal, we construct reciprocal space maps (RSMs) in the vicinity of the super-lattice reflection~\cite{Schlepuetz_2010,supplement}. 

\begin{figure*}
\includegraphics[width=1\textwidth]{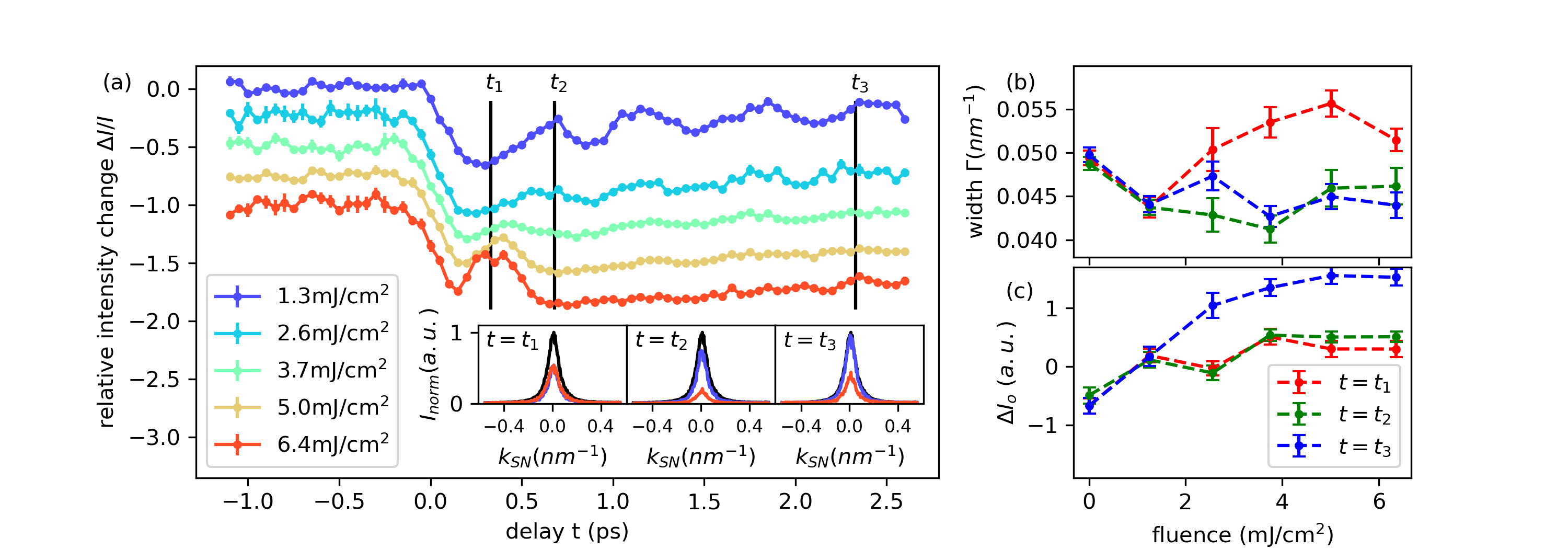}
\caption{\label{fig:Exp} Time-resolved diffraction data from the 
\((3, 7.252, -2.5)\)  reflection. (a) Transient evolution of the diffracted x-ray intensity for various absorbed fluences (vertically shifted for clarity). The black vertical lines indicate the three delays $t_1, t_2, t_3$ at which the RSMs were constructed. The insets show the RSM along the surface normal at each of the three delays for the unexcited sample (black) and at $f_\text{pump}=1.3 \si{\milli\joule}/\si{\square \centi\metre}$ (blue) and $f_\text{pump}=6.4 \si{\milli\joule}/\si{\square \centi\metre}$ (red). (b) Width of the fit of Eq.~\ref{eq_Lorentzian} to the projection of the RSM along the surface normal as function of the fluence $f_\text{pump}$ for the three delays. (c) Offset of the \(\Delta I_o\) extracted from the fit for the three delays. In both plots (b) and (c), the fluence $f_\text{pump}=0 \si{\milli\joule}/\si{\square \centi\metre}$ corresponds to times when the x-ray pulse arrives before the pump.}
\end{figure*}

The transient changes of the diffracted x-ray intensity for the optimal diffraction geometry are plotted in Fig.~\ref{fig:Exp}(a) for various absorbed fluences $f_\text{pump}$. At the lowest fluence ($f_\text{pump}=1.3 \si{\milli\joule}/\si{\square \centi\metre}$), the intensity of the scattered x-rays drops within $0.2 \si{\pico\second}$ followed by oscillations with a frequency of approximately $1.7 \si{\tera\hertz}$. At later delay times it relaxes to a level similar to that observed before the excitation. At higher fluences, the amplitude of the oscillation decreases and the average scattered x-ray intensity is suppressed over the entire range of investigated pump-probe delays. At the two highest fluences ($f_\text{pump} = 5.0$ and $6.4 \si{\milli\joule}/\si{\square \centi\metre}$) we observe a partial recovery of the super-lattice diffraction peak at about $t_1 = 0.33 \si{\pico\second}$ which coincides in delay time with the first minimum of the oscillation observed at the lowest fluence. 

To obtain a complete picture of the transient structure at different depths of the sample, we performed phi-scans (azimuthal rotation of the sample) at the delays $t_1 = 0.33 \si{\pico\second}$, $t_2 = 0.68\si{\pico\second}$ (one full oscillation period after time overlap of pump and probe at low fluence) and $t_3 = 2.33\si{\pico\second}$. The projection of the RSM constructed from these measurements along the surface normal (2, 0, -1) is plotted for selected delays and absorbed fluences in the insets of Fig.~\ref{fig:Exp}(a). The projections are each fit to a Lorentzian lineshape 
\begin{eqnarray}\label{eq_Lorentzian}
I(k) &= A \frac{\Gamma}{\pi(\Gamma^2 + (k-k_0)^2)} + \Delta I_o
\end{eqnarray}
where $k$ is the projection of the momentum along the surface normal, and $k_0$, $A$, and $\Gamma$ are fit parameters for the peak center, amplitude, and width (half-width-at-half-maximum), respectively. The additional fit parameter \(\Delta I_o\) represents an overall momentum-independent offset, which could include background contributions from fluorescence or diffuse scattering. The fitted values of $\Gamma$ and $\Delta I_o$ are shown in Fig.~\ref{fig:Exp}(b) and (c) respectively. Note that the unexcited diffraction peak has a width $\Gamma = 0.049\si{\per\nano\meter}$ corresponding to a coherence length of $l_c = 20\si{\nano\meter}$. This is approximately a factor of ten smaller than the probing depth of $210 \si{\nano\meter}$. We also see a marked broadening of the RSM at $t_1$ for fluences above $f_\text{pump}=3 \si{\milli\joule}/\si{\square \centi\metre}$. For later delays ($t_2$ and $t_3$) the width becomes narrower. The offset $\Delta I_o$ shows a significant increase at $t_3$ for fluences above $f_\text{pump}=2 \si{\milli\joule}/\si{\square \centi\metre}$ relative to $t_1$ and $t_2$.

\begin{figure}[b]
\includegraphics[width=0.45\textwidth]{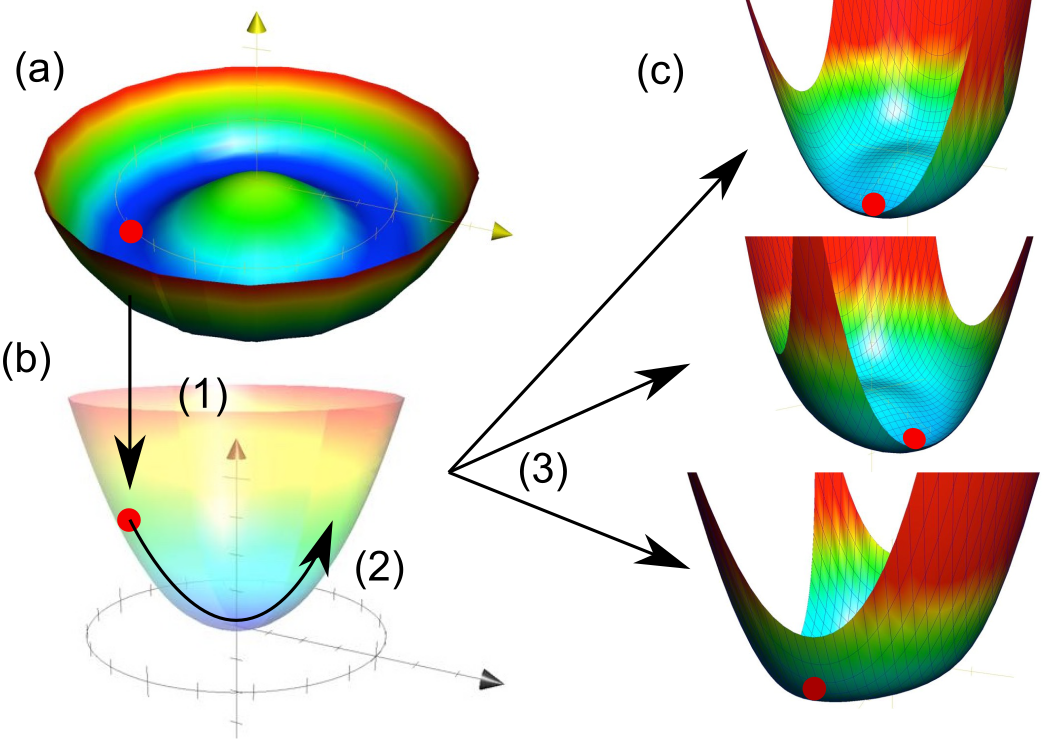}
\caption{\label{fig:Pot} Sketch of possible order-parameter dynamics: vertical axis represents the free energy, the radial distance the amplitude of the CDW and the angle (with respect to the x-axis) the phase. (a) Initial ``sombrero'' potential where the order parameter is denoted by a red dot. 
A strong excitation (1) turns the sombrero potential into a single well potential (b), where the order parameter moves through $\Psi =0$ and overshoots (2). The potential relaxes within an oscillation period (3), but defects along different chains favor different phases (c) such that the order parameter loses long-range coherence.}
\end{figure}

We interpret these observations in terms of a Ginzburg-Landau formalism adapted for dynamics of CDW systems~\cite{Yusupov_2010,Trigo_2021}. The CDW is characterized by a time- and space-dependent order parameter $\Psi\in \mathbb{C}$ with  free energy 
\begin{eqnarray}\label{eq_GLEn}
F_\text{GL} = \frac{a}{2}|\Psi|^2 + \frac{b}{4} |\Psi|^4 + \xi^2\left|\frac{\partial \Psi}{\partial z}\right|^2
\end{eqnarray}
where $z$ is the depth, $a = (T-T_c)/T_c$, $b>0$ and $\xi$ is the coherence length. At temperatures $T<T_c$, the energy landscape takes the form of the ``sombrero'' potential shown in Fig. \ref{fig:Pot}. 
The photoexcitation is modeled as a time-dependent electronic temperature 
\begin{equation}
T(t, z) = T_0+\eta\Theta(t) e^{-t/\tau}e^{-z/\delta}\label{eq_Temp}
\end{equation}
where \(T_0 < T_c\) is the initial temperature, $\eta$ is the peak surface temperature increase, \(\Theta\) is the Heaviside function, \(\tau\) is a relaxation time, and \(\delta\) is the intensity attenuation length of the pump laser. Assuming a constant electronic specific heat \(C_v^{(e)}\), \(\eta = \mathcal{F}_\mathrm{abs}/C_v^{(e)}\delta \),  where \(\mathcal{F}_\textrm{abs}\) is the absorbed laser fluence.   At sufficiently high fluences where \(\eta > T_c-T_0\), the parameter \(a\) becomes positive for a short time just after the excitation near the surface where the pump pulse enters. This makes \(\Psi = 0\) the only stable equilibrium value of the order parameter. Since the initial value of the order parameter is non-zero (such that \(|\Psi| = \sqrt{-a/b}\)), this causes a coherent oscillation of the order parameter around \(\Psi = 0\) \cite{Beaud_2014}. This can lead to a transient inversion of the initial complex phase of \(\Psi\) as the order parameter ``overshoots'' the zero-valued quasi-equilibrium point. Previous x-ray experiments \cite{Huber_2014, Neugebauer_2019} observed a partial transient recovery of super-lattice diffraction intensity that was interpreted as such an overshoot. The transient recovery seen in Fig.~\ref{fig:Exp}(a) for $5.0$ and $6.4 \si{\milli\joule}/\si{\square \centi\metre}$ at time $t_1$ can also be explained in this way. 

The surface-normal projections of the RSMs at particular delay times provide more information about the CDW order dynamics. At the lowest fluence ($f_\text{pump}=1.3 \si{\milli\joule}/\si{\square \centi\metre}$) the projection slightly narrows after excitation (see FIG.~\ref{fig:Exp}(b)). 
This narrowing may be caused by the depth-dependence of the sample excitation:  near the surface, where the CDW domain size is smaller   \cite{supplement}, the transient electronic temperature increase is higher and so these smaller domains are preferentially suppressed.
At absorbed fluences above ($3 \si{\milli\joule}/\si{\square \centi\metre}$) we observe a strong broadening (in increase of about 10\%) at the delay $t_1$. This observation is consistent with the expected effects of dynamic CDW phase inversion, which introduces additional domain walls at a depth set by the changes in the excitation level with distance into the sample.

The subsequent narrowing of the peak at $t_2$ and $t_3$ strongly suggests that this phase inversion vanishes after only one cycle of oscillation, as also visible in the transient evolution of the x-ray intensity at $6.4 \si{\milli\joule}/\si{\square \centi\metre}$ (Fig. ~\ref{fig:Exp}(a)). One explanation could be that the system remains in a single well potential and the order parameter quickly relaxes to a value of zero. As discussed in Ref.~\cite{Huber_2014, Neugebauer_2019} this would require a strongly time-dependent damping of the order parameter to explain why the oscillations fail to persist beyond 600~fs. An alternative explanation is that the electronic temperature \(T\) quickly relaxes to below \(T_c\), but this relaxation is accompanied by a depth-dependent randomization of the phase of the CDW. We propose that this randomization of the phase is driven by a high concentration of strong pinning defects in the crystal that locally favor particular phases of the CDW.  These defects are nucleation centers that result in initially very small domains and weak diffuse scattering, which may also explain the observed increase in the offset \(\Delta I_o\) observed at time \(t_3\).  Due to fairly weak interchain interactions the reestablishment of the original CDW domain structure takes much more time, similar to that observed in the formation of incommensurate CDW order in  TaS\(_2\)~\cite{Laulhe_2017, Jarnac_2021}.

\begin{figure*}
\includegraphics[width=0.95\textwidth]{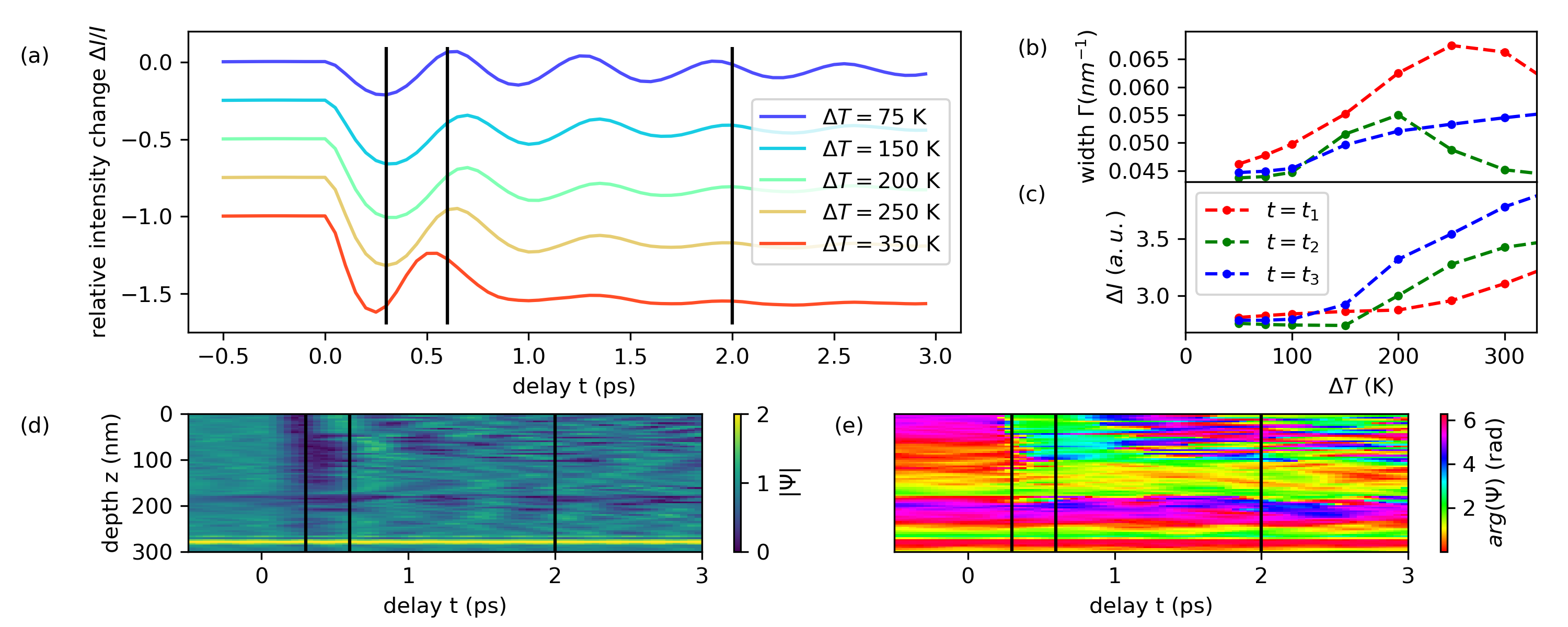}
\caption{\label{fig:Sim} Simulated transient diffraction data, using the model described in the text. (a) The transient evolution of the diffracted x-ray intensity for various excitations $\Delta T$. The black vertical lines indicate the three delays $t_1, t_2, t_3$ at which the RSMs were evaluated. (b) The width of the fit of the RSM projection as function of the excitation levels $\Delta T$. (c) The offset of the RSM for the three delays. (d) and (e) show the absolute value and phase of the order parameter as function of the depth below the surface ($z=0$) and delay for the excitation $\Delta T = 250\si{\kelvin}$ at an arbitrary in-plane position. The initial condition is taken from a thermalization process \cite{supplement} 
Laser excitation occurs at \(t=0\). The phase inversion close to the surface is visible in (d) as a dark line at $t_1$ and as a change of color from violet/red to green in (e). 
}
\end{figure*}

To investigate this further, we perform simulations where we numerically solve the GL equation \cite{Huber_2014, Neugebauer_2019, Trigo_2021} modified to include pinning defects \cite{supplement,Bixon_1971, Chen_2006, PenaMunoz_2023}. 
We treat the complex order parameter \(\Psi(t,x,y,z)\) in three spatial dimensions and assume a random, spatially uniform distribution of pinning defects with density \(n\)  at sites \(\vec{r}_i\), each favoring a random phase \(\phi_i\) of the CDW order. 
We also include in the model interaction with a phononic heat bath with temperature \(T_l\) that is coupled to the electronic temperature via a two-temperature model. The excitation is parametrized by the increase in quasi-equilibrium surface temperature $\Delta T$  at long delay times. As a result the time dependence of our electronic temperature does not follow the simple model of Eq.~\ref{eq_Temp}.
From the numerically solved order parameter $\Psi(t, x,y,z)$, the expected x-ray diffraction pattern is calculated and visualized in Fig.~\ref{fig:Sim}.

 Both the transient evolution of the integrated peak intensity (Fig.~\ref{fig:Sim} (a)) and fit parameters of the RSM projection (Fig.~\ref{fig:Sim} (b) and (c)) show qualitative agreement with the experiment. The simulation reproduces the suppression of the diffracted x-ray intensity at higher excitations and later delays (Fig.~\ref{fig:Sim}(a)) as well as the significantly larger increase of intensity in the RSM projection width at high excitations (Fig.~\ref{fig:Sim}(b)) and the large increase in diffuse background diffraction at $t_3$ (Fig.~\ref{fig:Sim}(c)). Looking at the evolution of the order parameter as a function of depth and time (see Fig.~\ref{fig:Sim}(d) and (e)), the inversion of the phase around $t_1$ is visible as well as the very fast dephasing around $0.8 \si{\pico \second}$ and the subsequent disordered configuration. 

 Under the assumption that the absorbed energy density \(\rho\) locally thermalizes at sufficiently long delay times and neglecting transport effects such as heat diffusion, it relates to \(\Delta T\), the transient increase in temperature, as $\rho = \int_{T_0}^{T_0+\Delta T} C(T) \,\mathrm{d}T$ where $C(T)$ is the heat capacity \cite{Kwok_1991}. Accordingly, the five laser fluences in the experiment each correspond to an increase in temperature $\Delta T $ of $80 \si{\kelvin}$, $130 \si{\kelvin}$, $170 \si{\kelvin}$, $210 \si{\kelvin}$, and $250 \si{\kelvin}$, respectively.

The qualitative discrepancies between the model and the experimental results are mostly visible in the RSM fitting parameters (Fig.~\ref{fig:Sim}(b) and (c) compared to Fig. \ref{fig:Exp}(b) and (c)). We believe these differences may be a result of several factors. First, in our actual sample the CDW is more disordered (i.e. has smaller domains and more pinning defects) close to the surface. This was not taken into account in the simulation as the depth dependence of the defect density is unknown. Second, the simulation employs a finite momentum grid to balance computational feasibility and accuracy; the complex dynamics we are simulating required us to limit the extent of momentum space to keep the computation time reasonable, therefore their resolution is lower than that of the experiment and differences in peak-to-background contrast are anticipated. In particular the finite RSM range impacts the offset $\Delta I_o$. 

Another disagreement between simulation and experiment concerns the transient evolution of the super-lattice diffraction intensity (Fig.~\ref{fig:Sim} (a)). In contrast to the experimental data in Fig.~\ref{fig:Exp}, the time \(t_1\), where the first minimum of the intensity is seen at low excitation levels, does not coincide with the time when the first local maximum of the intensity is seen in the high-excitation-level simulation. This discrepancy arises from the model's assumption that the excitation can be treated as a transient modification of the parameter \(a \sim T - T_c\), which governs the softening of the frequency of the order parameter oscillations. However,  this softening is not observed experimentally~\cite{pouget_1989, Schaefer_2010, Thomson_2017, Travaglini_1983}. As elaborated by Schaefer et al. \cite{Schaefer_2014} a simulation with two coupled order parameters, i.e. an electronic and a lattice (extendable to several lattice order parameters), reveal a suppressed mode softening in case the electronic order cannot adiabatically follow the lattice. The non-softening of the phonon mode explains the experimentally observed peak at high fluences coinciding with the minimum at low fluences: At low excitations, where the potential stays in the broken symmetry, the order parameter has a minimal value after half the oscillation whereas in case of high excitations, where the order parameter moves in a single well potential, the minimal value of the order parameter is reached after a quarter of an oscillation and a (local) maximum after half an oscillation. The experimentally observed shift of the partial recovery of the super-lattice diffraction towards shorter delays at higher fluences originates from probing the dynamics within the broken and high-symmetry phases at shallower and deeper depths correspondingly. The higher the fluence, the more pronounced the contribution from phase-inverted layers and the earlier the peak appears.

By performing ultrafast x-ray diffraction and reconstructing RSMs at selected delays we have shown that at excitations where a phase inversion occurs, the dynamics of the CDW after the phase inversion is dominated by decoherence of the CDW. We attribute this decoherence to pinning defects in the sample and the high anisotropy that hinders screening. Our suggestion of defects as a source of disorder is also consistent with the results of  previous double optical pump and time resolved x-ray probe experiments \cite{Neugebauer_2019}, where not only one recovery of x-ray intensity was observed but two. This indicates that by applying two pumps, the system can be kept in the single well potential for a longer time enabling for a coherent back oscillation. This suggests that the laser excitation itself does not cause decoherence. Simulations extending the GL model including these pinnings as well as coupling to an excited heat bath reproduce the key features observed in the experiment. Previous experiments describing similar observations with a time-dependent damping constant \cite{Huber_2014, Neugebauer_2019} were also able to reproduce the dynamics, but this approach was purely phenomenological and lacked a microscopic explanation. Our proposed model closes this gap by taking into account the theoretically and experimentally expected static impact of defects \cite{Fukuyama_1978, DeLand_1991} on the CDW phase and their role on the dynamics of the CDW after photoexcitation. 

Our results indicate that contributions from spatial disorder are critical to a complete description of ultrafast dynamics of the CDW in \BB in a high excitation regime. A similarly critical role of disorder has been suggested for the laser-driven dynamics of VO\(_2\) near its phase transition~\cite{Wall_2018}. Our results contrast with reports of persistent phase inversion in other materials such as SmTe\(_3\)~\cite{Trigo_2021} where the dynamics were modeled using a one-dimensional model without defects or consideration of a heat bath. One possible reason for the importance of these factors for \BB is the relatively low coherence length \(l_c\) (and correspondingly high density of pinning defects) along with a high anisotropy that hinders defect screening. An interesting possible avenue for future work would be to controllably change the density of pinning defects in a single material type to identify the crossover between these behaviors.

\begin{acknowledgments}
We thank P. Beaud and M. Trigo for fruitful discussions.
We acknowledge the Paul Scherrer Institut, Villigen,
Switzerland for provision of free electron laser beam-
time at the Bernina instrument of the SwissFEL Aramis
branch and synchrotron radiation beamtime at the AD-
DAMS beamline of the Swiss Light Source.
This research was supported by the Swiss National Science Foundation through Grant 192337.
A.N. acknowledges the Marie Skłodowska-Curie Grant Agreement No. 884104 (PSI-FELLOW-III-3i). 
This research was funded by the Schweizerischer Nationalfonds zur Förderung der Wissenschaftlichen Forschung
through Ambizione Grant PZ00P2\_179691.
\end{acknowledgments}

The data that support the findings of this article are openly available at https://doi.org/10.3929/ethz-c-000791063.

\bibliography{Bibliography_main}

\let\oldbib\bibliography          
\renewcommand{\bibliography}[1]{} 

\include{supplement}

\let\bibliography\oldbib          

\end{document}

%% file: Supplement.tex
\setcounter{equation}{0}
\setcounter{figure}{0}

\clearpage

\onecolumngrid
\begin{center}
\textbf{\large Supplementary Information:  Ultrafast Decoherence of Charge Density Wave Phase in K$_{0.3}$MoO$_3$}
\vspace{0.5 cm} 

{Rafael T. Winkler$^{1, 4}$, Larissa Boie$^1$, Yunpei Deng$^2$, Matteo Savoini$^1$, Serhane Zerdane$^2$, Abhishek Nag$^2$, Sabina Gurung$^1$, Davide Soranzio$^1$, Tim Suter$^1$, Vladimir Ovuka$^1$, Janine Zemp$^1$, Elsa Abreu$^1$, Simone Biasco$^1$, Roman Mankowsky$^2$, Edwin J. Divall$^2$, Alexander R. Oggenfuss$^2$, Mathias Sander$^2$, Christopher Arrell$^2$, Danylo Babich$^2$, Henrik T. Lemke$^2$, Urs Staub$^2$, Jure Demsar$^3$, Steven L. Johnson$^{1,2}$ } \\

\vspace{0.2 cm}

{$^1$Institute for Quantum Electronics, Physics Department, ETH Zurich, Zurich, Switzerland. \\
$^2$Center for Photon Science, Paul Scherrer Institute, Villigen, Switzerland. \\
$^3$Institute of Physics, Johannes Gutenberg-University Mainz\\
$^4$Institute of Science and Technology Austria (ISTA), Am Campus 1, 3400 Klosterneuburg, Austria
}

(Dated: \today)
\end{center}
\vspace{0.5 cm}


\section{Experimental Setup}

The x-ray experiment was performed at the Bernina end station at SwissFEL, PSI \cite{Prat_2020, Ingold_2019, Mankowsky_2021}. The sample was a single crystal with the surface cut parallel to the (2, 0, -1) planes. For the diffraction experiment it was cooled to $40\si{\kelvin}$ in a helium cryostat and positioned such that the surface normal was in the horizontal plane. The angle of incidence of the x-rays was set to $0.73^\circ$ and a spot size of $10\si{\micro\meter}$ x $125\si{\micro\meter}$ (at normal incidence), such that the beam spot was larger vertically. The photon energy of the x-rays was $7.5 \si{\kilo\electronvolt}$, set by the Si(111) monochromator.

The sample was excited with pulses of duration of $35 \si{\femto\second}$ at a wavelength of $510 \si{\nano\meter}$. The excitation pulses were generated from an amplified Ti:Sapphire laser system by mixing the signal output from an optical parametric amplifier with the 800 nm wavelength fundamental. The spot size of the focus was $330\si{\micro\meter}$ x $350\si{\micro\meter}$ at the sample location (at normal incidence), and had a grazing incidence angle of $13^\circ\pm 2^\circ$. To maximize the absorbed energy, the polarization was chosen to be horizontal. 

To take into account the temporal jitter between the x-ray and the laser, the standard spectral-encoding time tool of the Bernina endstation was used to correct for the jitter \cite{Ingold_2019, Bionta_2011}.

\section{X-Ray Diffraction Experiments}

This section reports some more details on the time resolved x-ray experiments. First details on the evaluation of the delay scans are given, then a description on the reciprocal space maps (RSMs) including their fitting. In the end results from RSMs at different angles of incidence are presented investigating the depth dependence of the coherence length.

\subsection{Delay Scans}

\begin{figure}
\includegraphics[width= 0.6\textwidth]{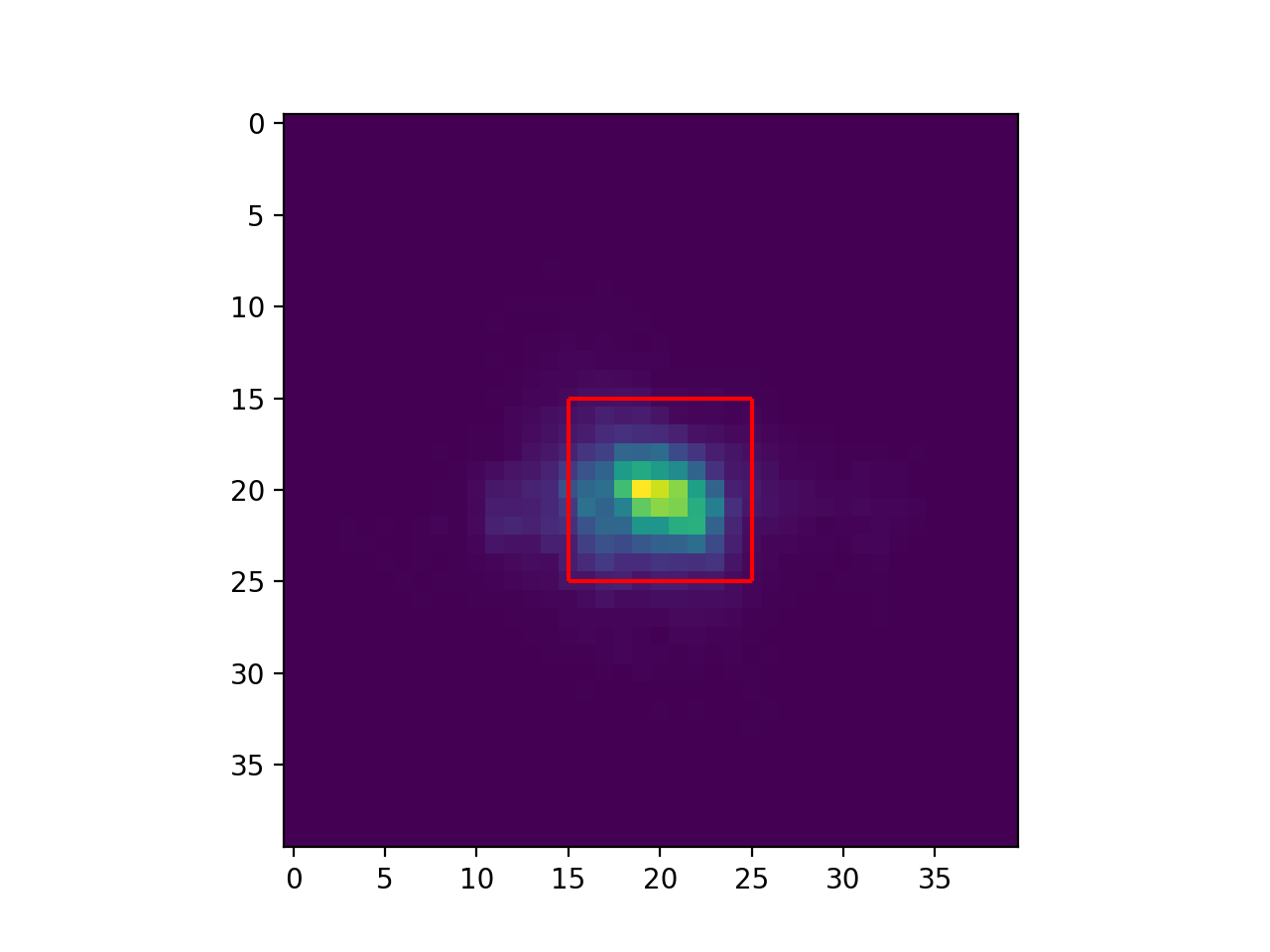}%
\caption{\label{Fig_Det_Delay} Image of the detector at the optimized scattering condition. The red square (10x10 pixels) denotes the region that was integrated for the delay scans.}
\end{figure}

The diffraction condition was optimized by rotating the sample around its azimuthal angle. The delay scans (i.e. Fig.~1(a) in the main manuscript) were done only at these optimized scattering conditions. Figure~\ref{Fig_Det_Delay} shows an image of the detector (Jungfrau \cite{Redford_2016}) with the diffraction peak and the integration region, resulting in an integrated intensity $I^{(i)}$ for each shot $i$. 

To account for the intensity jitter of the x-rays, for each shot the intensity of the x-rays before the sample was recorded, denoted by $I_0^{(i)}$. The shots were then time-tool corrected and binned. For each such bin $j$, the integrated intensity from the detector was then averaged over the corresponding shots and then normalized by the average reference intensity
\begin{align}\label{eq_I0norm}
  I_\text{tot}^{(j)} = \frac{\sum_{i} I^{(i)}}{\sum_{i} I_0^{(i)}}
\end{align}
where the index $i$ runs over all shots assigned to the bin \(j\).

\subsection{Reciprocal Space Map Reconstruction}

To calculate the RSM \cite{Schlepuetz_2010}, for each pixel of the detector (Jungfrau module) and azimuthal angle (phi), the scattered momentum transfer $\vec{q} = (h,k,l)^T-(h_0, k_0, l_0)^T$ (with T = transpose) relative to the center of the peak $\vec{q}_0 = (h_0, k_0, l_0)^T$ was calculated. Then, each of these scattering vectors was mapped onto the orthonormal system with $\vec{Q} = (k_x, k_y, k_z)^T$. This momentum transfer $\vec{Q}$ has units of inverse nanometres. Formally, this linear map is given by 
\begin{align}
\vec{Q} = M \cdot \vec{q}
\end{align}
where $M = \left(\vec{a^\ast}\left|\vec{b^\ast}\right|\vec{c^\ast}\right)$ is the matrix with columns equal to the reciprocal lattice vectors $\vec{a^\ast}$, $\vec{b^\ast}$ and $\vec{c^\ast}$. 

Since the surface normal of \BB is perpendicular to (2, 0, -1), the canonical basis vectors of the reciprocal space do not align with the surface normal which is the direction of interest in our paper. In order to project the reciprocal space maps (RSM) along the directions perpendicular to the surface normal, the reciprocal space map coordinates were rotated before performing the binning. In this way, one axis of the bins was aligned with the surface normal and the other axis perpendicular to the surface normal such that the projection is a summation of these two axes. The rotation is given by the map 
\begin{align}
\vec{Q}_R &= R \cdot\vec{Q}\\
R &= \begin{bmatrix}
(2,0,-1)\cdot M^T/||(2,0,-1)\cdot M^T||\\
(0,1,0)\cdot M^T/||(0,1,0)\cdot M^T|| \\
\left((2,0,-1)\cdot M^T\times (0,1,0)\cdot M^T\right)/||(2,0,-1)\cdot M^T\times (0,1,0)\cdot M^T||
\end{bmatrix}
\end{align}
with $\vec{Q}_R$ the rotated vector in reciprocal space. This map rotates the surface normal onto the first basis vector, the crystallographic $b$ axis onto the second basis vector and the direction normal to both onto the third basis vector. 

The detector orientation and resolution lead to a very uneven distribution of the number of pixels entering the different RSM bins. To prevent aliasing effects arising from this, we mimicked an increased resolution by interpolating the intensity between different pixels on the detector. This way, we obtained an interpolated dataset with three times the density of pixels along each detector direction, i.e. the final number of data points was nine times larger than the original one. We then performed the binning on this interpolated finer pixel grid and calculated the average normalized intensity in each bin $I(\vec{k})/I_0$, where \(I_0\) is the corresponding measured \(I_0\) signal.  Note that we lack the information needed to calibrate the values of the intensities to account for the actual scattering efficiently, so we report the values of \(I(\vec{k})/I_0\) as arbitrary (e.g. see Fig.~1(c) in the main text).

\begin{figure}
\includegraphics[width= 0.9\textwidth]{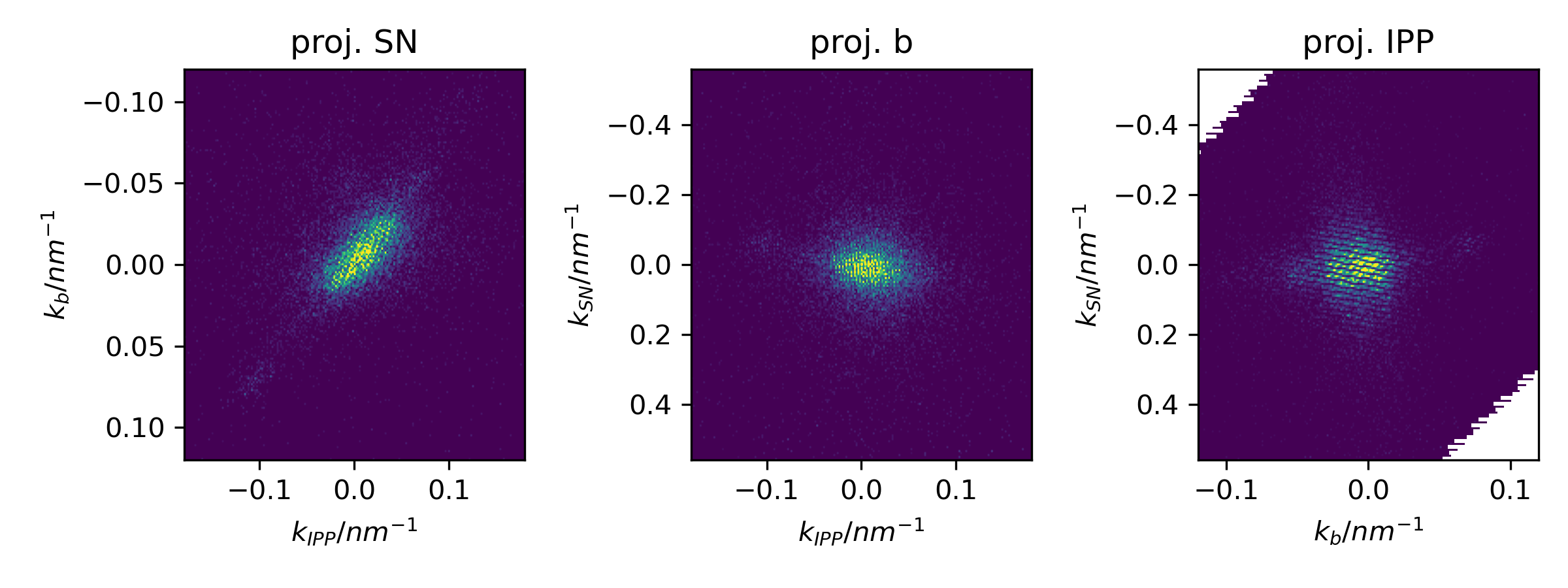}%
\caption{\label{Fig_RSM_Proj} RSMs projected along the three axes: (a) along the surface normal, (b) along $b$ and (c) along the direction perpendicular to the two (IPP) at $t_1$ and the highest pump fluence $6.4 \si{\milli\joule\per\square \centi\metre}$. Note that the axes along the different directions have a different scales.}
\vspace{-86mm} \hspace{0mm} \textbf{(a)} \hspace{40mm} \textbf{(b)} \hspace{40mm} \textbf{(c)} \hspace{30mm}
\vspace{81mm}
\end{figure}

The projection of the rotated RSM along each basis vector is visualized in Fig.~\ref{Fig_RSM_Proj}. We observed an elongation of about a factor 3 along the surface normal (i.e. Fig.~\ref{Fig_RSM_Proj}(b) and (c)) indicating a shorter coherence along the surface normal. Considering the projected RSM along the surface normal (Fig.~\ref{Fig_RSM_Proj}(a)) there is another, smaller elongation visible. The orientation of that elongation matched the azimuthal angle of the sample for various diffraction peaks. Therefore, we identify this effect to originate from the elongated footprint of the x-rays on the sample.

\subsection{Fitting of Projected RSMs}

To reveal the broadening and background of the RSMs at the different fluences and delays, the projected RSMs along the surface normal were fitted by a Lorentzian lineshape 
\begin{eqnarray}\label{eq_Lorentzian_S}
I(k) &= A \frac{\Gamma}{\pi(\Gamma^2 + (k-k_0)^2)} + \Delta I_o
\end{eqnarray}
where $k$ is the projection of the momentum along the surface normal, and $k_0$, $A$, $\Gamma$ and $\Delta I_o$ are fit parameters for the peak centre, area, width (HWHM) and offset respectively.

Although the Lorentzian lineshape fits the data well, the offset $\Delta I_o$ takes slightly negative values for the unexcited sample (as visible in Fig.~1(c) in the main text). This suggests that the long tails characteristic of the Lorentzian may not perfectly describe the actual lineshape of the RSM projection. Indeed the covariance matrix of the fit shows a negative correlation between the area $A$ and the offset $\Delta I_o$ as well as between the width $\Gamma$ and $\Delta I_o$. This effect cannot explain the observed increase in $\Delta I_o$ at the latest delay $t_3$ compared to $t_2$, as the scattered intensity increases from $t_2$ to $t_3$ and $\Gamma$ does not significantly change.
Due to the relatively small magnitude of the unexcited \(\Delta I_o\) we did not judge it useful to model the peak with a more complex functional shape.

\subsection{Depth-Dependent Defect Concentration}

\begin{figure}
\includegraphics[width= 0.9\textwidth]{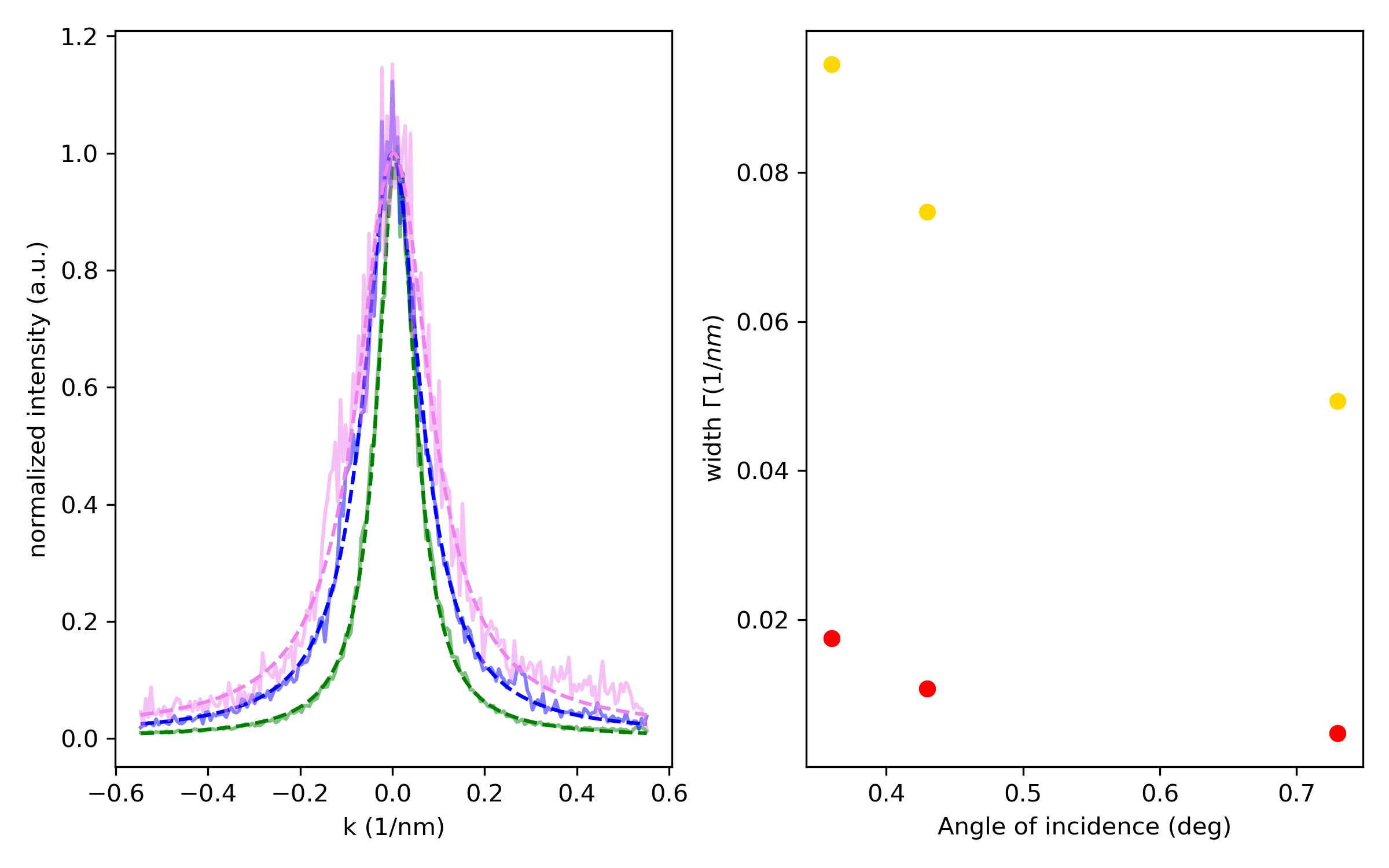}%
\caption{\label{Fig_RSM_AOI} Dependence of the angle of incidence  on the width of the projection of the RSM along the surface normal direction. (a) Normalized RSM projection (solid lines) along the surface normal for the three incidence angles $0.73^\circ$ (green), $0.43^\circ$ (blue) and $0.36^\circ$ (violet) and the fit as dashed line. The traces are normalized by the maximum of the fit. (b) Comparison of the width for different angle of incidence. The yellow dots represent the width $\Gamma$ from the fits (a). The red dots are the expected broadening $\Gamma_d= 1/\delta$ due to the finite penetration depth $\delta$.}
\vspace{-150mm} \hspace{-20mm} \textbf{(a)} \hspace{70mm} \textbf{(b)} \hspace{44mm}
\vspace{148mm}
\end{figure}

To investigate the depth dependence of the defect concentration, we measured the RSMs as a function of the angle of incidence of the x-rays.  Because of the grazing incidence geometry, a decrease in the incidence angle measures a region of the sample closer to the surface. Fig.~\ref{Fig_RSM_AOI} shows the RSMs of the unexcited sample for three different incidence angles ($0.73^\circ$, $0.43^\circ$ and $0.36^\circ$, projected along the surface normal direction. Comparing the width at the largest angle of incidence with the smallest, one can see that it increases from $0.048\si{\per\nano\metre}$ to $0.098\si{\per\nano\metre}$, i.e. doubles. Part of this broadening can be explained by the shorter penetration depth of the x-rays at shallower angles. However, comparing again the largest and smallest angles, the width increases by only  $0.013\si{\per\nano\metre}$ \cite{Henke_1993}. This implies that the shorter penetration depth cannot explain the increase of the measured width. We therefore attribute this increase of width to a shorter coherence length of the CDW close to the surface originating from a larger defect concentration.

\section{Simulation}

Several previous computational models \cite{Huber_2014, Yusupov_2010, Neugebauer_2019, Trigo_2021, Picano_2023} of the ultrafast dynamics of the CDW order describe the order parameter $\Psi$ as function of depth $z$ in a double well potential of the form 
\begin{align}
V(\Psi) &= \int-\frac{c}{2}|\Psi(\vec{r})|^2 + \frac{1}{4}|\Psi(\vec{r})|^4 \mathrm{d} V,
\end{align}
where the phase transition is triggered by a change in the prefactor $c$ due to the photoexcitation of the laser taking the form $c(t, z) = 1-\eta e^{-t/\tau}e^{-z/\delta}$ where $\tau$ is the relaxation time of the excitation and $\delta$ the penetration depth.

As described in the main text, in order to model our data we have extended this model in several respects:
\begin{itemize}
\item We include three spatial dimensions to take into account for interactions between chains and the corresponding anisotropy of the sample.
\item We also consider impurities pinning the phase of the CDW, causing incoherent phase modes after the excitation.
\item We introduce an effective electronic and lattice temperature that are connected via a two-temperature model \cite{Chen_2001, Chen_2006}.
\end{itemize}
The different aspects and their implementation are discussed in the subsequent sections. The last two sections summarize the resulting equations of motion and the procedure to obtain the parameters in the simulations.

\subsection{Increase of Dimensionality and Frame of Simulation}\label{tit_Dim_Sim}

The high anisotropy of the sample originating from the chain-like structure prevents a screening of defects and consequently enhances the coherence of the CDW within the MoO$_3$ sheets compared to the coherence between the sheets. To account for this, the order parameter at each simulation site is a complex number $\Psi(t_i, \vec{r})\in \mathbb{C}, \vec{r} = (x_j, y_k, z_l)$. The order parameter is complex to account for different phases across the sample.

To account for the spatial anisotropy, a gradient term is included in the potential, which then takes the form
\begin{align*}\label{Pot_ext}
V(\Psi) = \int &-\frac{c}{2} \Psi(\vec{r})\Psi(\vec{r})^\ast + \frac{d}{4}(\Psi(\vec{r})\Psi(\vec{r})^\ast)^2 \\
&+ \frac{\xi_z^2}{2}\left|\partial_z \Psi(\vec{r})\right|^2 + \frac{\xi_\perp^2}{2}\left|\vec{\nabla}_\perp \cdot \Psi(\vec{r})\right|^2\mathrm{d} V 
\end{align*}
where $\xi_z$ and $\xi_\perp$ denote the coherence strength along the surface normal and perpendicular to it respectively. Similarly the spatial derivatives are given by $\partial_z$ and $\vec{\nabla}_\perp$. This treatment is motivated by our RSM data, where we see a similar peak width in the two in-plane directions.

\subsection{Impurities and Defects Pinning the Phase}\label{tit_Impurities}

The electronic contribution to the CDW makes it sensitive to defects and impurities which locally modify the electronic structure. The impurities are included in the simulation by deformations of the potential at positions $\vec{r}_j$, where $j =1, 2, ...N$. The positions of these deformations are distributed uniformly and randomly with density $\rho= N/V_s$, where $V_s$ is the volume of the simulation.   The deformations are modeled by adding a term
\begin{align*}
V_p(\Psi) &=-\sum_{j} Re(k^\ast\delta(\vec{r}-\vec{r}_j)\Psi(\vec{r})) |\Psi(\vec{r})|^{2}\\
&= -\sum_{j} \delta(\vec{r}-\vec{r}_j)|k| \cos(\varphi-\varphi_j) |\Psi(\vec{r})|^{3}
\end{align*}
to the potential,
where $k\in\mathbb{C}$ and $\varphi_j$ denote the strength and phase of the pinning located at site $j$ and $\varphi$ is the phase of the CDW at $\vec{r}_j$. The potential is assumed to scale with the third power of $|\Psi|$ which was chosen because the second power and fourth power are already present in the potential and the first power rather corresponds to the interaction of the order parameter with an external field. 

The full potential that is used for the simulations is then
\begin{equation}
\begin{split}
V(\Psi) = \int \Bigg[&-\frac{c}{2} \Psi(\vec{r})\Psi(\vec{r})^\ast + \frac{d}{4}(\Psi(\vec{r})\Psi(\vec{r})^\ast)^2 
+ \frac{\xi_z^2}{2}\left|\partial_z \Psi(\vec{r})\right|^2 + \frac{\xi_\perp^2}{2}\left|\vec{\nabla}_\perp \cdot \Psi(\vec{r})\right|^2\\ 
&-\sum_{j} \delta(\vec{r}-\vec{r}_j)|k| \cos(\varphi-\varphi_j) |\Psi(\vec{r})|^{3} \Bigg]dV
\end{split}\label{Pot_ext}
\end{equation}

\subsection{Two Temperature Model and Heatbath}\label{Sim_2T}

In our model we use a two-temperature model to approximate the changes to the electronic states and the general vibrational modes of the lattice as a function of time after excitation \cite{Chen_2001, Chen_2006}.
We assume that the pump laser interacts predominantly via a sudden increase in electronic temperature $T_e(t, z)$, which both triggers the order parameter dynamics and eventually relaxes via a transfer of heat to the lattice. The lattice temperature $T_l(t, z)$ acts as a heat bath for the order parameter dynamics. Because of the finite attenuation length of the pump, both \(T_e(t,z)\) and \(T_l(t,z)\) are time- and depth-dependent. We define $T_f(z)$ as the common final temperature of the electrons and lattice at long times (but at times sufficiently short that we can ignore heat diffusion). The temporal evolution of the two temperatures is thus given by 
\begin{align}
T_e(t, z) &= G(t) \ast (T_f-T_0) \Theta(t) \left(M_e e^{-\frac{t}{\tau}}+1\right)e^{-z/\delta}+T_0\\
T_l(t, z) &=(T_f-T_0) \Theta(t) \left(1-e^{-\frac{t}{\tau}} \right)e^{-z/\delta}+T_0 
\end{align}
where \(\Theta\) is the Heaviside function and $G$ is a Gaussian kernel with FWHM $100\si{\femto\second}$, taking into account the finite duration of the laser and x-ray pulse. This Gaussian kernel is then convoluted with the actual function where $M_e$ denotes the ratio between phononic and electronic heat capacities causing the electronic temperature to exceed the final temperature on short time scales. $T_0$ represents the initial temperature. The strength of the excitation is parametrized by the long-term increase in temperature $\Delta T = T_f-T_0$ which is related to the absorbed laser fluence by the heat capacity of the sample \cite{Kwok_1991}.

In our model we assume that the parameter c in equation (\ref{Pot_ext}) depends on the electronic temperature \(T_e\) as 
\begin{align*}
c(T_e) = \frac{T_c^{(e)}-T_e}{T_c^{(e)}}.
\end{align*}
where $T_c^{(e)}$ denotes the critical temperature where the Sombrero potential changes to a single well potential. 

The lattice temperature is used for two different aspects of the model. To model the effects of the elevated lattice temperature on the CDW coherence length, we make the spatial coherence parameter $\xi$ temperature dependent 
\begin{equation}
\label{eq_chi_T}
\xi(T_l) = \begin{cases}
    \xi_0 & T_l <T_c^{(l)}\\
    0 & T_l \geq T_c^{(l)}\\
\end{cases}.
\end{equation}
where $T_c^{(l)}$ is the transition temperature where the coherence is lost. We also include the lattice temperature in a Langevin term which was added to the equation of motion. This Langevin term is related to the damping constant and introduces random fluctuations with a vanishing mean but a variance that scales linearly with the temperature \cite{Bixon_1971, PenaMunoz_2023}.

\subsection{Equation of Motion}

Attributing a mass $m$ to the order parameter and correspondingly a kinetic energy, one can derive the equation of motion according to the Euler-Lagrange equation. By renormalizing the equation such that the Sombrero potential has prefactors of $\Psi$ and $|\Psi|^2\Psi$ equal to $1$, the equation of motion reads as
\begin{align}\label{eq_EOM}
0 &= \frac{2}{\omega_{AM}^2} \frac{\partial^2 \Psi}{\partial t^2} + \gamma\frac{\partial \Psi}{\partial t} - \xi_z^2 \frac{\partial^2 \Psi}{\partial z^2} - \xi_\perp^2\left(\frac{\partial^2 \Psi}{\partial x^2} + \frac{\partial^2 \Psi}{\partial y^2}\right) - \Psi + \Psi |\Psi|^2 \nonumber\\ 
&- \sum_{j}\left(k\delta(\vec{r}-\vec{r}_j) |\Psi(\vec{r})|^2 - 2Re(k^\ast\delta(\vec{r}-\vec{r}_j) \Psi(\vec{r}))\Psi(\vec{r})\right)-\eta(T_l).
\end{align}
where $\omega_{AM}^2= c/m$ is the angular frequency of the amplitude mode, $\gamma$ is the damping constant and $ \eta(T_l)$ the Langevin term. The other variables are described in the previous sections.

\subsection{Selection of Parameters}

Table \ref{Tab_Par} gives an overview of the parameters and their values used in the simulations. Some parameters could be directly obtained by corresponding references or fits, others had to be calculated indirectly, in particular:
\begin{itemize}
  \item The coherence parameter $\xi$ is related to the frequency of the phase mode as 
  \begin{align}
    \nu_p = \sqrt{\frac{\xi^2 \omega_\text{AM}}{2}}. 
  \end{align}
  The phase mode frequency \(\nu_p\)was estimated by Pouget et al. \cite{Pouget_1991}. Furthermore, from the RSM we observe a three times broader diffraction peak along the surface normal than in the in-plane direction. Hence $\xi_\perp=3\xi_z$.
  \item The density of defects $\rho$ and their pinning strength $|k|$ (for a given choice of $\xi_z$ and $\xi_\perp$) determine the width of the RSM in the absence of excitation. The chosen values showed the best agreement between simulation and experiment. Furthermore, the density $\rho$ agrees with a study by DeLand et al. \cite{DeLand_1991} where they investigated the broadening of the RSM as function of displaced Mo atoms. At a broadening similar to our case, they estimate the density of displaced Mo atoms to be $\rho_\text{Mo} \approx 10^{-2}\si{\per\cubic\nano\meter}$. However, to agree with theory, they argue that only one out of thousand displaced Mo atoms actually leads to a pinning defect.
  \item The laser fluence required for the change in symmetry of the potential depends not only on $T_c^{(e)}$ but also on the ratio of the phononic and electronic heat capacities $M_e$, i.e. most relevant is the ratio $M_e/T_c^{(e)}$. Therefore, we set $T_c^{(e)}$ and adapted $M_e$ accordingly.
  \item In the renormalized equation \ref{eq_EOM}, the temperature at which the Langevin term drives the order parameter over the energy barrier is also dependent on the mass $m$. With all other parameters set, the mass is the only free parameter. The transition sets in at about $m=1.6 \cdot10^{-22}\si{\kilogram}$, to keep the temperature dependence of the coherence term the main contribution (as given by equation \ref{eq_chi_T}) of the phase transition, the mass was increased by a factor of 4 to $m = 6.2 \cdot10^{-22}\si{\kilogram}$.
\end{itemize}

As anticipated in section \ref{Sim_2T}, the excitation is parametrized by the final temperature $\Delta T = T_f-T_i$. The five laser fluences in the experiment correspond to increase in temperature of $\Delta T = 80, 130, 170, 210, 250 \si{\kelvin}$ respectively.

\begin{table}[h]
\caption{Parameters used in the simulation. \label{Tab_Par}}
\begin{ruledtabular}
\begin{tabular}{lll}
Parameter & value & Comment \\
\hline
$\omega_{AM}$ & $2\pi \times 1.7\si{\tera\hertz}$& \cite{Schaefer_2010}\\

$\gamma$&$0.006\si{\pico\second}$& \cite{Yusupov_2010}\\

$\xi_z$&$2.4\si{\nano\meter}$& \cite{Pouget_1991}, see also  text\\

$\xi_\perp$&$3\xi_z$& comparison RSM, see also  text\\

$|k|$&$4000$& comparison RSM, see also  text\\

$\rho$&$5\cdot 10^{-5}\si{\per\cubic\nano\meter}$& comparison RSM, see also  text\\

$m$&$6.2 \cdot10^{-22}\si{\kilogram}$& corresponds to a phase transition at about $600\si{\kelvin}$\\

$T_c^{(e)}$&$1000\si{\kelvin}$& guess in combination with $M_e$, see also  text\\

$T_c^{(l, z)}$&$180\si{\kelvin}$& phase transition, see also  text \\

$T_c^{(l, \perp)}$&$300\si{\kelvin}$& phase transition, see also  text \\

$\tau$&$0.2\si{\pico\second}$& \cite{Huber_2014} \\

$\delta$&$175 \si{\nano\meter}$& \cite{Sagar_2007}\\

$M_e$&5&agreement simulation, see also  text\\
\end{tabular}
\end{ruledtabular}
\end{table}

\subsection{Initial Conditions}

The initial conditions for the simulation are determined by additional simulations that approximate the order parameter in thermal equilibrium.
 They proceed as follows:
\begin{enumerate}
    \item In the first step, the electronic temperature \(T_e\) is set to \(4 T_c^{(e)}/3\) to obtain a single-well potential, but the coherence parameters $\xi_\perp$ and $\xi_z$ are set to the values given in Table~\ref{Tab_Par}. The Langevin term is set to zero and the damping \(\Gamma\) set to $0.04\si{\pico\second}$.  The order parameter is set to zero everywhere  except at the pinning sites where it is fixed to the pinning value.
    \item Eq.~\ref{eq_EOM} is numerically integrated to solve for the dynamics of the order parameter over a time span of two picoseconds.  This results in non-zero values of the order parameter near the pinning sites.
    \item Over a further time interval of $10\si{\pico\second}$ of numerical integration of Eq.~\ref{eq_EOM}, the electronic temperature \(T_e\) is set to decrease linearly to $0\si{\kelvin}$. 
    \item Eq.~\ref{eq_EOM} is integrated  for another $38\si{\pico\second}$ with $T_e = 0\si{\kelvin}$.  This is intended to remove any residual dynamics of the order parameter.
    \item At this point in the simulation the Langevin term is re-introduced in the equation of motion and the damping set to the value in Table~\ref{Tab_Par}.  With these new parameters Eq.~\ref{eq_EOM} is integrated for an additional two picoseconds at \(T_e = 0\si{\kelvin}\).
    \item Eq.~\ref{eq_EOM} is then integrated for 8 ps as the electronic temperature \(T_e\) increases linearly to \(40\si{\kelvin}\).
    \item The simulation then proceeds for an additional 40 ps with \(T_e = 40\si{\kelvin}\).
\end{enumerate}
The end result of this process is used as a set of initial conditions for the simulations of the pump-probe dynamics.

This procedure was performed for various values of pinning densities $\rho$ and pinning strength $|k|$. By evaluating the width of the computed projection of the RSM for different combinations at $0\si{\kelvin}$, we chose five pairs with very different values that are consistent with the experimentally observed RSM width to perform simulations with photoexcitation. Comparing the RSM projections after photoexcitation of the simulations with the experiment, we found a similar behavior of the background $\Delta I_0$ for all configurations. Comparing the five sets of pinning paramers, there are some differences in the general transient evolution of the simulated scattered x-ray intensity as well as in the values of the width at $t_1$, $t_2$ and $t_3$ which could also be related to the softening of the phonon mode close to the phase transition. We present  in Table~\ref{Tab_Par} the combination of $\rho$ and $|k|$ that fits the experiment best.

\newpage

\bibliography{Bibliography_supplement}